\documentclass[aip,cha,reprint]{revtex4-1}
\usepackage{amscd}
\usepackage{amssymb}
\usepackage{amsfonts}
\usepackage{amsmath}
\usepackage{graphicx}
\usepackage{dcolumn}
\usepackage{bm}
\usepackage{multirow}
\usepackage[dvipdfm]{hyperref}

\begin{document}

%\preprint{AIP/123-QED}

\title{Optimizing and controlling functions of complex networks by manipulating rich-club connections}
\author{Xiao-Ke Xu}
%\email{xuxk@eie.polyu.edu.hk}
\affiliation{Department of Electronic and Information Engineering,
Hong Kong Polytechnic University, Hong Kong, People's Republic of
China}
\affiliation{School of Communication and Electronic
Engineering, Qingdao Technological University, Qingdao 266520,
People's Republic of China}

\author{Jie Zhang}
\affiliation{Department of Electronic and Information Engineering,
Hong Kong Polytechnic University, Hong Kong, People's Republic of
China} \affiliation{Centre for Computational Systems Biology, Fudan
University, Shanghai 200433, People's Republic of China}

\author{Ping Li}
\affiliation{College of Informatics, South China Agricultural
University, Guangzhou 510642, People's Republic of China}

\author{Michael Small}
\affiliation{Department of Electronic and Information Engineering,
Hong Kong Polytechnic University, Hong Kong, People's Republic of
China}

\date{\today}

\begin{abstract}
Traditionally, there is no evidence suggesting that there are strong
ties between the rich-club property and the function of complex
networks. In this study, we find that whether a very small portion
of rich nodes connected to each other or not can strongly affect the
frequency of occurrence of basic building blocks (motif) within
networks, and therefore the function, of a heterogeneous network.
Conversely whether a homogeneous network has a rich-club property or
not generally has no significant effect on its structure and
function. These findings open the possibility to optimize and
control the function of complex networks by manipulating rich-club
connections. Furthermore, based on the subgraph ratio profile, we
develop a more rigorous approach to judge whether a network has a
rich-club or not. The new method does not calculate how many links
there are among rich nodes but depends on how the links among rich
nodes can affect the overall structure as well as function of a
given network. These results can also help us to understand the
evolution of dynamical networks and design new models for
characterizing real-world networks.
\end{abstract}

\pacs{89.75.Fb, 89.75.Hc, 89.75.Da}

\maketitle

\begin{quotation}
The function properties of complex networks --- properties such as
synchronizabilty, and efficiency of information transport --- depend
sensitively on the detailed topological structure of the particular
network. We show that, despite this, the functional behavior of a
complex network can be largely controlled by rewiring a very small
fraction of nodes within the network. For networks with scale-free
degree distribution (heterogeneous networks) the connectivity
pattern among the highest degree nodes determines the functional
behavior of the entire network. That is, whether a network exhibits
a dominant rich-club (whether high degree nodes are mutually
connected or not) will determine the functional behavior of the
entire network. For random graphs and other networks with
homogeneous degree distribution this is not the case. Our results
provide a mechanism by which the behavior of real-world networks can
be effectively controlled by rewiring only a very small portion of
links. This suggests the likely mechanism by which real-world
networks such as the internet and gene regulatory networks in
various organisms evolve. The connectivity among a small portion of
hub nodes will control the functional behavior of the entire
network, and thus a real network can be made more robust or provide
more efficient information processing by rewiring those links.
\end{quotation}

\section{Introduction}
%motif and the function of complex networks
The motif, defined as a small connected subgraph that recurs in a
graph, is the basic building block, or functional unit, of complex
networks \cite{motif_block}. In real-world networks (e.g., gene
regulatory networks), motifs represent the elementary interaction
patterns between small groups of nodes, and the relative frequencies
with which motifs appear represent different functions of the
network \cite{Motif, Review_motif, Superfamily}. Although it has
been found that there is a topological relationship between the
large-scale attributes (scale-free and hierarchical) and local
interaction patterns (subgraph based) \cite{motif_pnas}, it remains
unclear whether there is a relationship between small functional
units and other structure properties such as rich-club connections
of complex networks. In our previous study we find that rich-club
connections can dominate some global properties (e.g., assortativity
and transitivity) of a network \cite{Xu_unpublished}, which implies
the possible relation between the rich-club property and the
network's subgraph organization.

%The study on rich-club
The rich-club property refers to the organization pattern of rich
nodes \cite{Richnode_PA}, especially whether rich nodes tend to
connect to one another, or with the remaining nodes
\cite{Richclub_origin, Colizza_richclub, NP_comment, Constraint,
APL_richclub, Zhou_richclub}. Because rich nodes often play a
central role in the static property of, and dynamic processes on,
complex networks \cite{Attack, Cascading_failure, Super_spreader},
significant attention has been paid to the prominent effects of the
richest elements \cite{AS_Xu} and the organization among them
\cite{Weight_richclub, Xu_unpublished}. A systematic framework is
needed to clearly understand the roles of rich nodes in different
real-world networks with distinct degree distributions.

In this study, we find the influences of rich nodes and their
organization pattern depend largely on the degree distributions of
complex networks. Rich nodes are important in scale-free networks
\cite{BA}, because a power-law degree distribution indicates that
the majority of nodes participate in at most one or two motifs,
while a few rich nodes take part in a very large number of small
subgraphs. Manipulating a very small number of rich-club connections
therefore can strongly affect the frequencies of the basic
functional blocks (motifs) for a heterogeneous network. In
comparison, for the network with a homogeneous degree distribution
(e.g., the network of US power grid), the links among rich nodes
show a tiny effect on the whole network. The main reason behind this
is that all nodes (including rich nodes) in a homogeneous network
are engaged in only a few interactions, and there are no hubs
linking to a significantly larger number of other nodes.

These results are helpful in understanding the origin of motifs and
motif clusters in real-world complex networks, and the mechanisms by
which how small subgraphs aggregate into larger superstructures. Our
finding has an important potential application: we can build a
framework to optimize and control the functional behaviors of
complex networks. In most cases we can not regenerate or redesign a
real-world network, but manipulating a small number of rich-club
connections gives us a chance to optimize the structure of the
network and control the relative frequencies of small functional
units in a predictable manner.

%what we have done
Furthermore, although pioneer studies have developed a series of
methods to judge whether a network has rich-club properties
\cite{Constraint, Colizza_richclub, Zhou_richclub}, these approaches
are based on how many links there are among rich nodes instead of
how these links affect the whole network. Based on subgraph ratio
profile, the topological structure among rich nodes can be uncovered
from the inspection of the basic functional units. In this study we
develop a novel method to judge whether a network has a rich-club or
not. The new method does not calculate how many links connect to
rich nodes compared with its randomized version while it depends on
how the organization pattern of rich nodes affects the appearance of
different motifs.

%The relationship of rich-club and motif.
Taken together, these findings indicate the strong ties between the
local subgraphs and rich-club properties of complex networks, which
complements our understanding of a network's topological and
functional organization. Because each network can be characterized
by a set of distinct types of subgraphs and rich-club connections
are a significant property, our findings are expected to provide new
insights in understanding the evolution of dynamical networks and
design new models for characterizing real-world networks. Our work
is a step in an ongoing effort to bridge the local topology of a
network and its global statistical features.

\section{Method}
\subsection{Link rewiring algorithms}

%%%%%%%%%%%%%%%%%%%%%%%%%%Figure 1 start%%%%%%%%%%%%%%%%%%%%%%%%%%%%%
%summary how to do it
Here we select the top $0.5\%$ of the nodes with the
highest degree as rich nodes in a network and manipulate
the connections among them. We use link rewiring algorithms
to generate the network with rich-club and the network
without rich-club, respectively. The basic idea is very
similar to the random rewiring method
\cite{Correlation_Science}, while the main difference is
that our new method only switches the links among rich
nodes and a small number of low-degree nodes. First we make
rich nodes fully connected to one another, so they form a
completely connected rich-club. Secondly, we completely
eradicate the edges among rich nodes, so that the network
has no rich-club.

\begin{figure}[htbp]
\includegraphics[width=0.48\textwidth]{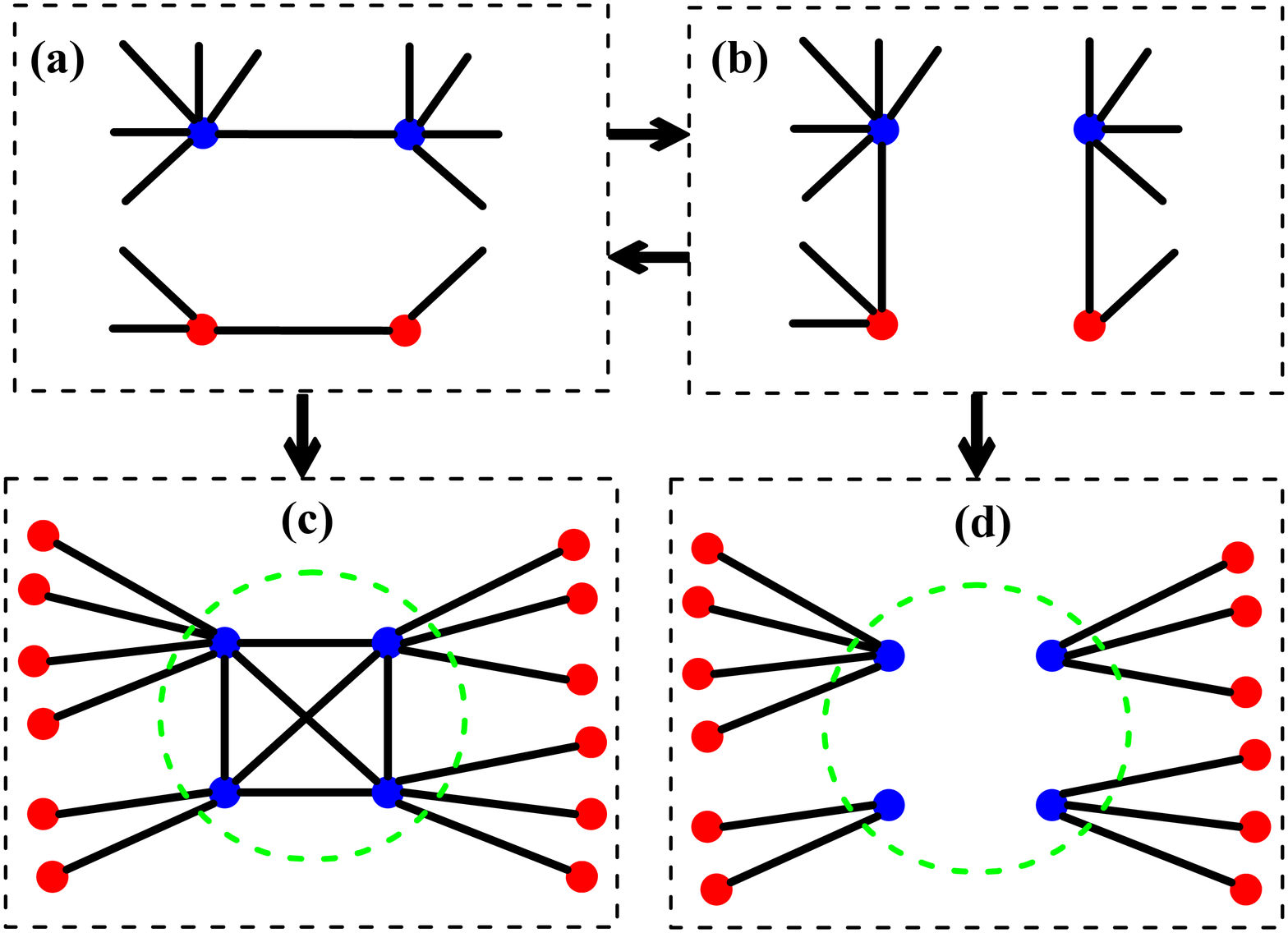}
\caption{(Color online) (a) and (b) are the two connection patterns
for the four end nodes of a pair of links. (a) rich-club connection,
where one link connects to the two rich nodes and the other link
connects to the other low-degree nodes; (b) non-rich-club
connection, where one link connects to one rich node and one
low-degree node, and the other link connects to the two remaining
nodes. Using the link rewiring algorithms, we can obtain (c) the
network with rich-club, or (d) the network without rich-club.}
\label{fig1}
\end{figure}

% to be a rich-club network
Now we specify the rewiring algorithms. First we make all rich nodes
fully connected to generate a network with a significant rich-club.
If there is a link between two rich nodes, their structure remains
unchanged [Fig. \ref{fig1}(a)]. If there is no link between two rich
nodes, we perform the operation from Fig. \ref{fig1}(b) to
\ref{fig1}(a). That is, we select another two low-degree nodes that
respectively connect to the two rich nodes while do not connect to
each other [Fig. \ref{fig1}(b)]. Then we cut the two links between
the rich nodes and their low-degree neighbors, and connect the two
rich nodes as well as the two low-degree nodes, respectively [Fig.
\ref{fig1}(a)]. After repeating this process until all rich nodes
form a completely connected rich-club, we can obtain the network
with a full-connected rich-club [Fig. \ref{fig1}(c)].

% to be a non-rich-club network
Secondly, we completely eradicate the edges among rich
nodes, so that the network has no rich-club property. If
there is no link between two rich nodes, we will do nothing
[Fig. \ref{fig1}(b)]. If there is a link between two rich
nodes, we do the operation from Fig. \ref{fig1}(a) to
\ref{fig1}(b). We randomly select another pair of
low-degree nodes which connect to each other while do not
connect to either of the two rich nodes [Fig.
\ref{fig1}(a)]. Then we cut off the links both between the
two low-degree nodes and between two rich nodes
respectively, and let each rich node connect to one
low-degree node [Fig. \ref{fig1}(b)]. Repeating the above
process until the links among the whole rich nodes are
completely eradicated, we will get a network without
rich-club property [Fig. \ref{fig1}(d)].

% the meaning of our method
%why we do it as follows
Because we use the rewiring method, the degree of every
node in the original network exactly remains unchanged. For
the topological structure of the original network, there is
only small variation induced by manipulating rich-club
connections, so we can monitor how the subgraph frequencies
are affected by the rich-club property. Furthermore, we can
compare the results of the subgraph ratio profile for the
original network, the network with rich-club, and the
network without rich-club, to make more reliable inference
of whether the original network has a rich-club property or
not.

%%%%%%%%%%%%%%%%%%%%%%%%%%Figure 1 end%%%%%%%%%%%%%%%%%%%%%%%%%%%%%

\subsection{Motif clusters of rich nodes in non-rich-club and rich-club networks}

%%%%%%%%%%%%%%%%%%%%%%%%%%Figure 2 start%%%%%%%%%%%%%%%%%%%%%%%%%%%%%
%the definition of motif
Each network will be scanned for all possible $n$-node subgraphs (we
choose $n=4$). In a network with a skewed degree distribution, rich
nodes have much higher degrees than the overwhelming majority, so
whether they connect to each other to form a rich-club will strongly
affect the frequencies of subgraphs. Actually, rich nodes can absorb
a very large number of subgraphs and form a motif cluster. For
example, the triangles may not distribute uniformly within a
scale-free network but tend to aggregate around the hubs, because a
node with $k$ links can carry up to $k^2$ triangles
\cite{motif_pnas}. The aggregation of motifs into motif clusters is
important, because it implies that the potential functional
properties of the large number of subgraphs also need to be
evaluated at the level of subgraph clusters instead of being
evaluated only at the level of a single subgraph.

\begin{figure}[htbp]
\includegraphics[width=0.48\textwidth]{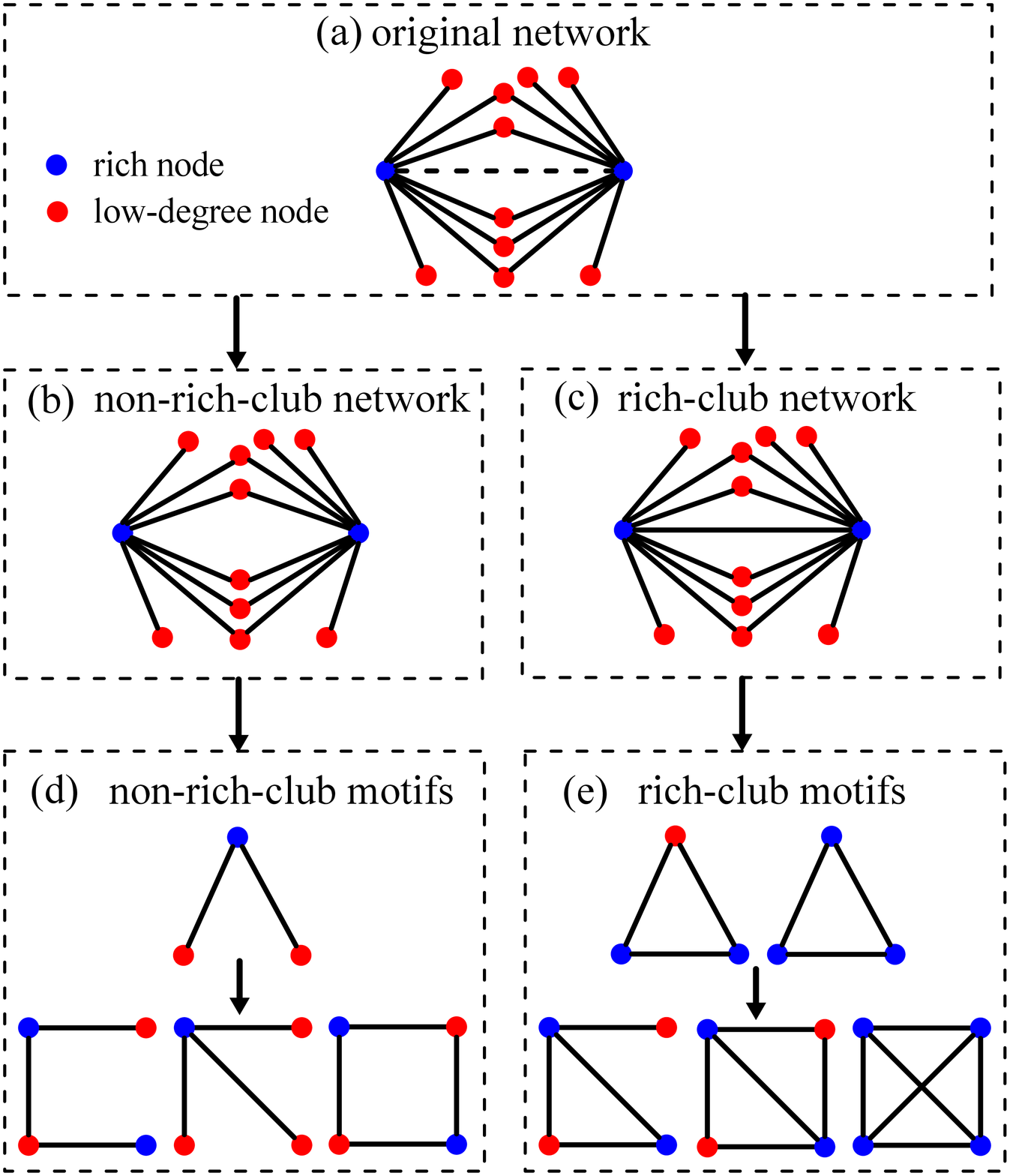}
\caption{(Color online) The demonstration for the aggregation of
non-rich-club motifs when a network has no rich-club and the
aggregation of rich-club motifs when a network has a rich-club.}
\label{fig2}
\end{figure}

Exploring rich-club connections provides a new way to evaluate the
functional properties of abundant subgraphs at the level of subgraph
clusters. A few rich nodes usually take part in a very large number
of small subgraphs and they can form motif clusters in real-world
complex networks. Actually, the organization of rich nodes can
dominate the appearance of particular motifs prominently. In the
non-rich-club network [Fig. \ref{fig2}(b)], rich nodes do not tend
to connect to each other, so the non-rich-club subgraphs [Fig.
\ref{fig2}(d)] will be more common. On the contrary, in the
rich-club network [Fig. \ref{fig2}(c)], rich nodes trend to connect
to each other, so the network will demonstrate a larger number of
the rich-club motifs [Fig. \ref{fig2}(e)]. In the original network
[Fig. \ref{fig2}(a)], rich nodes may or may not connect to each
other. Comparing the appearing frequencies of motifs in the above
three networks, we can conclude whether the original network has a
rich-club property.

It is obvious that by considering the subnetworks of rich nodes, the
frequencies of the non-rich-club motifs and/or rich-club motifs are
remarkably more than those of the randomized versions of the
subnetworks. The inherent existence of two distinct classes of
subgraphs (non-rich-club motifs and rich-club motifs) in a
heterogeneous network demonstrates that, in contrast to the
homogeneous network, the highly abundant motifs can not exist in
isolation but must naturally aggregate into subgraph clusters.
Specifically, in the network with a rich-club, the neighbors of a
highly connected node are linked to each other, therefore the chance
that low-degree nodes participate in highly connected subgraphs is
slim. In a homogeneous network, however, all nodes are engaged in
only a few interactions and the appearance of motifs is the
statistical average of the whole network, for there are no hubs
linking to a significantly higher number of other nodes to form
motif clusters.

%%%%%%%%%%%%%%%%%%%%%%%%%%Figure 2 end%%%%%%%%%%%%%%%%%%%%%%%%%%%%%

\section{Results}
\subsection{Motif distributions in homogeneous and heterogeneous networks}

%explanation for the value of $k_{max}/k_s$
Table \ref{table1} lists the results of six undirected networks
(including three real-world networks and three model networks)
arranged with $k_{max}/k_s$ increasing. The value of the structural
cutoff degree $k_s$ can be regarded as the first approximation of
the maximum degree within a scale-free network \cite{Degree_cutoff}.
Here $k_{max}/k_s$ is a convenient index that can be used in complex
networks with any degree distribution to show the proportion of
links (or degrees) rich nodes possess in comparison with the
remaining nodes in a network \cite{Xu_unpublished}.

%%%%%%%%%%%%%%%%%%%%%%%%%%Table 1 start%%%%%%%%%%%%%%%%%%%%%%%%%%%%%

\begin{table}[htbp]
\centering \caption{Statistics of six undirected networks: number of
nodes $n$, average degree $\langle k\rangle$, the exponent of degree
distribution if the distribution follows a power law: $\alpha$ (or
``--'' if not), structural cutoff degree $k_s=\sqrt{\langle k\rangle
n}$ \cite{Degree_cutoff}, maximal degree $k_{max}$. SW is the
network generated by the small-world model \cite{SW}, PG is the
network of US power grid \cite{BA}, BA is the network generated by
the scale-free model \cite{BA}, EPA is the network from the pages
linking to www.epa.gov \cite{Pajek_data}, PFP is the network
generated by the model for the Internet topology \cite{PFP} and AS
is the network of the Internet topology at the level of autonomous
systems \cite{ASdata}.}
\begin{ruledtabular}
\begin{tabular}{ c c c c c c c}
Network & SW & PG & BA & EPA & PFP & AS \\
\hline
$n$ & $5000$ & $4941$ & $5000$ & $4772$ & $5000$ & $5375$ \\
$\langle k\rangle$ & $6.0$ & $2.7$ & $6.0$ & $3.7$ & $6.0$ & $3.9$ \\
$\alpha$ & $-$ & $-$ & $3.0$ & $2.0$ & $2.2$ & $2.2$ \\
$k_{max}$ & $16$ & $19$ & $219$ & $175$ & $1259$ & $1193$ \\
$k_s$ & $173.2$ & $115.4$ & $173.2$ & $132.9$ & $173.2$ & $144.8$ \\
{$k_{max}/k_s$} & $0.09$ & $0.16$ & $1.26$ & $1.32$ & $7.26$ & $8.24$ \\
type & \multicolumn{2}{c}{$k_{max}\ll k_s$} & \multicolumn{2}{c}{$k_{max}\approx k_s$} & \multicolumn{2}{c}{$k_{max}\gg k_s$} \\
\end{tabular}
\end{ruledtabular}
\label{table1}
\end{table}

The low values of $k_{max}/k_s$ for SW and PG mean that the two
networks have a homogeneous degree distribution and the degrees of
rich nodes are close to the majority of nodes. While a high value of
$k_{max}/k_s$ indicates that the network has a heterogeneous degree
distribution and the degrees of a few rich nodes are far larger than
the rest, like BA and EPA. Especially, PFP and AS not only have a
power-law degree distribution, but also possess a few superrich
nodes \cite{AS_Xu} for $k_{max}\gg k_s$ in the two networks.

%%%%%%%%%%%%%%%%%%%%%%%%%%Table1 end%%%%%%%%%%%%%%%%%%%%%%%%%%%%%

%%%%%%%%%%%%%%%%%%%%%%%%%%Table 2 start%%%%%%%%%%%%%%%%%%%%%%%%%%%%%
%the properties of the two motifs
Although motifs are only local interaction patterns, the
distribution of motifs can greatly reflect the topological
properties of the networks \cite{motif_pnas, Timeseries_pnas}. In
Table \ref{table2}, we list the percentage of heterogeneous-motif,
the percentage of homogeneous-motif, and the percentage of the sum
of heterogeneous-motif and homogeneous-motif for all the networks.
The heterogeneous-motif is an unequal small structure: the blue
vertex represents a rich node, and the other three red vertexes
represent low-degree nodes. This non-equilibrium structure shows
that the three low-degree nodes all attach to the rich node, while
the low-degree nodes do not connect to each other. Obviously, the
rich node has the highest status in the four nodes and this
structure should appear more in a network with a heterogeneous
degree distribution. Especially, as in the case of many real-world
networks, subgraphs with a central node are abundant in a scale-free
network. The homogeneous-motif is a chain-structure, and the
stations of the four nodes are more likely to be equal. We assert
that this structure should frequently occur in a network with a
homogeneous degree distribution.

%the ratio increasing of the two motifs
As we have predicted, the results in Table \ref{table2} show that
the percentage of homogeneous-motif for SW [$61.9\%$] and PG
[$59.4\%$] are larger than the networks with a heterogeneous degree
distribution. The subset of $n$-node subgraphs in a heterogeneous
network often contains a central node, so the heterogeneous-motif
occurs more commonly in heterogeneous networks, such as for BA
[$64.5\%$], EPA [$80.5\%$], PFP [$92.8\%$] and AS [$96.0\%$]. In
summary, with the value of $k_{max}/k_s$ increasing, the ratio of
heterogeneous-motif/homogeneous-motif increases too.

\begin{table}[htbp]
\centering \caption{The first row is the percentage of
heterogeneous-motif, the second row is the percentage of
homogeneous-motif, the third row is the ratio of heterogeneous-motif
and homogeneous-motif, and the fourth row is the percentage of the
sum of the two subgraphs. SW, PG, BA, EPA, PFP and AS represent the
same networks in Table \ref{table1}.}
\begin{ruledtabular}
\begin{tabular}{ c | c c | c c | c c}
Motif & SW & PG & BA & EPA & PFP & AS \\
\hline
\includegraphics[height = 7 mm ]{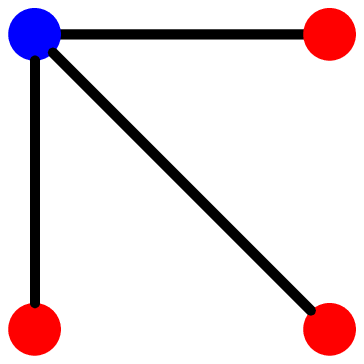} & $9.7\%$ & $31.3\%$ & $64.5\%$ & $80.5\%$ & $92.8\%$ & $96.0\%$ \\
\hline
\includegraphics[height = 7 mm ]{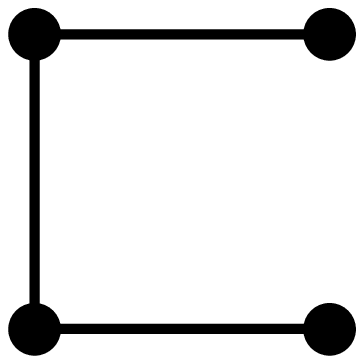} & $61.9\%$ & $59.4\%$ & $34.6\%$ & $18.5\%$ & $4.7\%$ & $3.1\%$ \\
\hline
\includegraphics[height = 7 mm ]{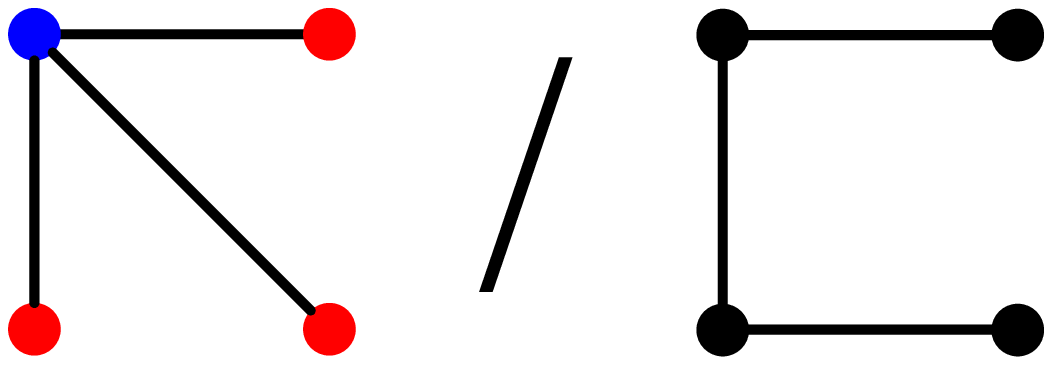} & $0.16$ & $0.53$ & $1.86$ & $4.35$ & $19.93$ & $31.40$ \\
\hline
\includegraphics[height = 7 mm ]{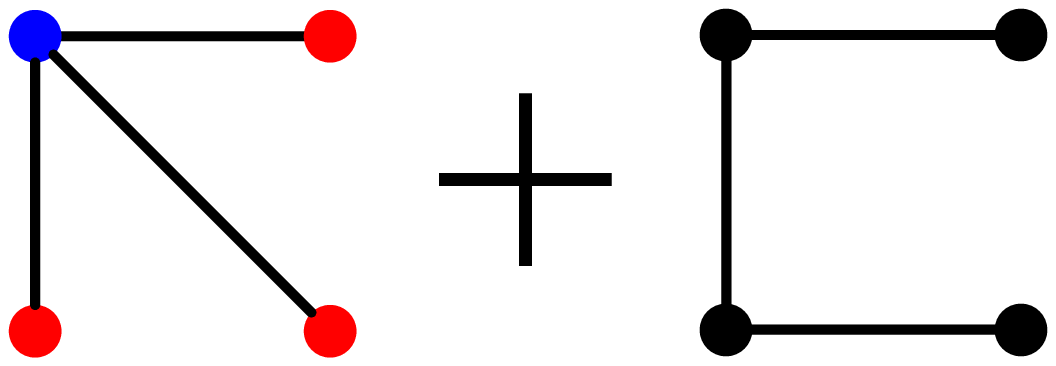} & $71.6\%$ & $90.7\%$ & $99.1\%$ & $99.0\%$ & $97.5\%$ & $99.1\%$ \\
\end{tabular}
\end{ruledtabular}
\label{table2}
\end{table}

The above results indicate that rich nodes in homogeneous networks
(e.g., SW and PG) only have a very limited effect on the whole
network, for all nodes (including rich nodes) in such type of
networks are engaged in only a few interactions. Obviously,
rich-club connections are more involved in the heterogeneous-motif
in a heterogeneous network, for a node with higher degree has a more
chance to participate in this structure. In a heterogeneous network
(e.g., BA, EPA, PFP and AS), a few rich nodes with much more links
than the overwhelming majority can absorb a very large number of
subgraphs and form motif clusters, which makes rich-club connections
more influential to the whole network.

The percentage of the sum of heterogeneous-motif and
homogeneous-motif for all the networks is very high (up to
$99.1\%$), which means other types of motifs are relative sparse
compared with the above two types of motifs. Therefore, to form a
specific functional block, the absolute frequencies of a particular
subgraph are not necessary very large. Actually, it is enough for
the relative frequencies of the motif for the original network are
statistically higher than those for its randomized version
\cite{Motif}. Moreover, in view of the difficulty in forming the
specific functional blocks in a randomized network, the sparse
distribution of other motifs gives us a chance to control the
appearance of small functional subgraphs in real-world networks by
manipulating rich-club connections.
%%%%%%%%%%%%%%%%%%%%%%%%%%Table 2 end%%%%%%%%%%%%%%%%%%%%%%%%%%%%%

\subsection{Superfamilies of non-rich-club and rich-club networks}
%the definition of SRP
Because undirected networks have only two types of triads (unclosed
triple and triangle), we only analyze the profile of the six types
of undirected connected tetrads ($4$-node motifs). The normalized
$Z$ scores of tetrads show a significant dependence on the network
size, so we use the abundance of each subgraph $i$ relative to
random networks \cite{Superfamily}:

\begin{equation}
\Delta_i=\frac{Nreal_i - \langle Nrand_i \rangle}{Nreal_i+ \langle
Nrand_i \rangle + \varepsilon}, \label{eq2}
\end{equation}
where $\varepsilon=4$ ensures that $\mid \Delta_i \mid $ is not
misleadingly large when the subgraph appears very few times in both
the real and random networks. The Subgraph Ratio Profile (SRP) is
the vector of $\mid \Delta_i \mid $ normalized to length 1:

\begin{equation}
SRP_i={\Delta_i}/{(\sum{{\Delta_i}^2})^{1/2}}. \label{eq3}
\end{equation}

%the introduction of Superfamily
Network motifs, which are patterns of interconnections occurring in
complex networks are significantly higher than those for randomized
networks \cite{Motif}. The motif pattern reflects the local
structural properties of complex networks and thus can be used to
classify networks. If different types of networks share the similar
result of SRP, these networks can be classified into the same
``superfamily'' \cite{Superfamily}. The networks in the same triad
superfamily share not only some particular types of motifs, but also
very similar proportions of all types of subgraphs.

\begin{figure}[htbp]
\includegraphics[width=0.48\textwidth]{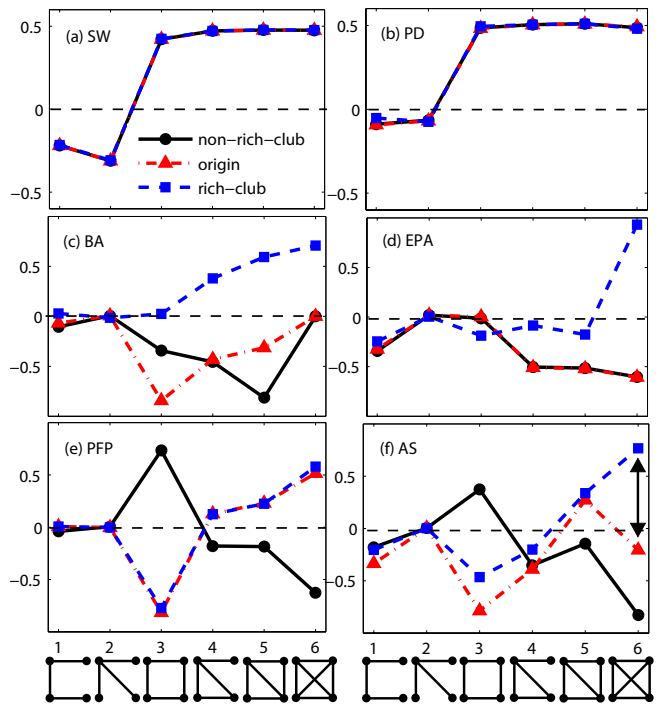}
\caption{(Color online) The subgraph ratio profile (SRP) for six
undirected networks. SW, PG, BA, EPA, PFP and AS represent the same
networks in Table \ref{table1}.} \label{fig3}
\end{figure}

%%%%%%%%%%%%%%%%%%%%%%%%%%Figure 3 start%%%%%%%%%%%%%%%%%%%%%%%%%%%%%
%the meaning of superfamily
Here we show the SRP results for the original network, the network
with rich-club, and the network without rich-club in Fig.
\ref{fig3}. If the above three networks belong to the same
superfamily, it means that the rich-club property has weak effect on
the original network, and this result shows the network is a
homogeneous network. If the three networks belong to different
superfamilies, it means that rich-club connections can strongly
affect the structure and function of the original network, and this
result indicates that the network is heterogeneous. Furthermore,
according to the fact that the original network belongs to the same
superfamily as the network with rich-club or the network without
rich-club, we can judge whether the original network has a rich-club
property.

%the results of SW and PG
The networks of SW and PG have a homogeneous degree distribution, so
rich nodes in the two networks are not significantly higher than
others. Therefore, as has been shown in Figs. \ref{fig3}(a) and
\ref{fig3}(b), whether the two networks have rich-club properties
does not have any influence on SRP. Moreover, the original network,
and the networks with and without rich-club all belong to the same
superfamily. The above results indicate whether a homogeneous
network has a rich-club property is not very important, and
rich-club connections can not control the functions of such type of
networks.

%the results of BA and EPA
Because the networks of BA and EPA have a heterogeneous degree
distribution, rich nodes possess much more links than the
overwhelming majority. Therefore, whether the two networks have
rich-club properties can greatly affect the result of SRP. As is
shown in Figs. \ref{fig3}(c) and \ref{fig3}(d), the original network
and the network with rich-club do not belong to the same
superfamily. Conversely, the original network and the network
without rich-club belong to the same superfamily, so BA and EPA both
have no rich-club property.

%the results of PFP and AS
For the networks of PFP and AS, they not only have a heterogeneous
degree distribution but also have a few superrich nodes, so whether
the two networks have a rich-club can affect the result of SRP most
significantly. The original network and the network without
rich-club do not belong to the same superfamily. For the original
PFP and the network with rich-club belong to the same superfamily,
PFP has the property of rich-club as is shown in Fig. \ref{fig3}(e).
Basically we can say that AS has a rich-club property, for the
original AS has the very similar SRP to the network with rich-club,
except for the motif $6$ ($4$-node clique) in Fig. \ref{fig3}(f).
Yet the non-consistency of motif $6$ for the original network and
the network with rich-club may be the origin of arguments on whether
the Internet topology has a rich-club property
\cite{Richclub_origin, Colizza_richclub, NP_comment, Constraint}.

%%%%%%%%%%%%%%%%%%%%%%%%%%Figure3 end%%%%%%%%%%%%%%%%%%%%%%%%%%%%%

\section{Conclusions}
%summary the findings
In conclusion, we find that the influences of rich-club connections
strongly depend on the degree distributions of complex networks. Our
findings show that in a homogeneous network, whether the network has
a rich-club or not is not very important for its structure and
function. While rich-club connections in a heterogeneous network
have a crucial implication, for they can partially optimize and
control the function of the whole network.

Our new framework for measuring the subgraph ratio profile can
provide a more impartial judgement on whether a network has a
rich-club. Previous studies put more attention on finding whether
the links among rich nodes appear more frequently in the original
network compared with its randomized counterparts
\cite{Colizza_richclub, NP_comment}. While the actual influence of
rich-clubs in different degree distribution networks has not been
studied. Our approach which is based on the effect of the rich-club
on the network structure and function, is therefore more advanced.

We demonstrate that strong ties between the rich-club property and
local (subgraph-based) structure underscore the importance to
understand the properties of complex networks as fully integrated
systems. Indeed, the abundance of some kinds of local interaction
patterns reflects the rich-club property of a network, raising
intriguing questions about the role of local events in shaping a
network's overall behavior \cite{motif_pnas}. These results indicate
that the analysis described here may have an impact on our
understanding for other types of subgraphs (e.g., cliques
\cite{Clique} and cycles \cite{Cycle}) in complex networks.

Our results show the significance of the rich-club property and
motif distributions in modeling and designing real-world networks
\cite{Network_motif_distribution}. An appropriate model should have
similar structure and function to the real-world network. To meet
this demand the model can be designed from the basic motifs or the
subgraph ratio profile, which can be easily controlled by the
rich-club property.

Our findings also deepen our understanding of the evolution of
dynamical networks. The existence of the dense rich-club motifs
and/or non-rich-club motifs in real-world networks may be a unifying
property of evolved systems, so it is interesting to understand the
rich-club concept from the perspective of network evolution. We
conjecture that the common origin of the local functional blocks and
the rich-club property is primarily the same, because neither the
density and topology of subgraphs nor the rich-club property can be
dissociated from the evolution of the overall network. Following the
framework in this work, we will contrive to bridge the gap between
local topologies of a network and its global statistical features in
future.

\begin{acknowledgments}
This work was supported by PolyU Postdoctoral Fellowships Scheme
(G-YX0N \& G-YX4A). X.-K. Xu and J. Zhang also acknowledge the
National Natural Science Foundation of China under Grant No.
61004104.
\end{acknowledgments}

%bibliography

\end{document}